\documentclass[twocolumn]{aastex62}

\newcommand{\kms}{km\,s$^{-1}$}
\newcommand{\HI}{H\,{{\small \sc I}}}
\newcommand{\CO}{CO($J$=0$\rightarrow$1)}
\newcommand{\CN}{CN($N$=0$\rightarrow$1)}
 
\graphicspath{{./}{figures/}}

\received{Initial Submission September 25, 2023}
\accepted{December 27, 2023}

\shorttitle{CO and CN absorption at z=3.4}
\shortauthors{Emonts et al.}

\begin{document}

\title{Absorption of Millimeter-band CO and CN in the Early Universe: Molecular Clouds in Radio Galaxy B2\,0902+34 at Redshift 3.4}

\correspondingauthor{Bjorn Emonts}
\email{bemonts@nrao.edu}

\author[0000-0003-2983-815X]{Bjorn H. C. Emonts}
\affiliation{National Radio Astronomy Observatory, 520 Edgemont Road, Charlottesville, VA 22903, USA}

\author[0000-0002-1163-010X]{Steve J. Curran}
\affiliation{School of Chemical and Physical Sciences, Victoria University of Wellington, PO Box 600, Wellington 6140, New Zealand}

\author[0000-0003-2884-7214]{George K. Miley}
\affiliation{Leiden Observatory, Leiden University, PO Box 9513, 2300 RA Leiden, The Netherlands}

\author[0000-0003-1939-5885]{Matthew D. Lehnert}
\affiliation{Universit\'{e} Lyon 1, ENS de Lyon, CNRS UMR5574, Centre de Recherche Astrophysique de Lyon, F-69230 Saint-Genis-Laval, France}

\author[0000-0001-6647-3861]{Chris L. Carilli}
\affiliation{National Radio Astronomy Observatory, P.O. Box O, Socorro, NM 87801, USA}

\author[0000-0001-9163-0064]{Ilsang Yoon}
\affiliation{National Radio Astronomy Observatory, 520 Edgemont Road, Charlottesville, VA 22903, USA}

\author[0000-0002-9482-6844]{Raffaella Morganti}
\affiliation{ASTRON, the Netherlands Institute for Radio Astronomy, Oude Hoogeveensedijk 4, 7991 PD Dwingeloo, The Netherlands.}
\affiliation{Kapteyn Astronomical Institute, University of Groningen, P.O. Box 800, 9700 AV Groningen, The Netherlands}

\author[0000-0002-0587-1660]{Reinout J. van Weeren}
\affiliation{Leiden Observatory, Leiden University, PO Box 9513, 2300 RA Leiden, The Netherlands}

\author[0000-0003-2566-2126]{Montserrat Villar-Mart\'{i}n}
\affiliation{Centro de Astrobiolog\'{i}a, CSIC-INTA, Ctra. de Torrej\'{o}n a Ajalvir, km 4, 28850 Torrej\'{o}n de Ardoz, Madrid, Spain}

\author[0000-0002-2421-1350]{Pierre Guillard}
\affiliation{Sorbonne Universit\'{e}, CNRS UMR 7095, Institut d'Astrophysique de Paris, 98bis bvd Arago, 75014, Paris, France}

\author[0009-0003-5974-0185]{Cristina M. Cordun}
\affiliation{ASTRON, the Netherlands Institute for Radio Astronomy, Oude Hoogeveensedijk 4, 7991 PD Dwingeloo, The Netherlands.}
\affiliation{Kapteyn Astronomical Institute, University of Groningen, P.O. Box 800, 9700 AV Groningen, The Netherlands}

\author[0000-0002-0616-6971]{Tom A. Oosteroo}
\affiliation{ASTRON, the Netherlands Institute for Radio Astronomy, Oude Hoogeveensedijk 4, 7991 PD Dwingeloo, The Netherlands.}
\affiliation{Kapteyn Astronomical Institute, University of Groningen, P.O. Box 800, 9700 AV Groningen, The Netherlands}



\begin{abstract}
Using the Karl G. Jansky Very Large Array (VLA), we have detected absorption lines due to carbon-monoxide, \CO, and the cyano radical, \CN, associated with radio galaxy B2\,0902+34 at redshift $z$\,=\,3.4. The detection of millimeter-band absorption observed 1.5 Gyr after the Big Bang facilitates studying molecular clouds down to gas masses inaccessible to emission-line observations. The CO absorption in B2\,0902+34 has a peak optical depth of $\tau$\,$\ge$\,8.6$\%$ and consists of two components, one of which has the same redshift as previously detected 21-cm absorption of neutral hydrogen (\HI) gas. Each CO component traces an integrated H$_{2}$ column density of $N_{\rm H_2}$\,$\ga$\,3\,$\times$\,10$^{20}$ cm$^{-2}$. CN absorption is detected for both CO components, as well as for a blueshifted component not detected in CO, with CO/CN line ratios ranging from $\la$0.4 to 2.4. We discuss the scenario that the absorption components originate from collections of small and dense molecular clouds that are embedded in a region with more diffuse gas and high turbulence, possibly within the influence of the central Active Galactic Nucleus or starburst region. The degree of reddening in B2 0902+34, with a rest-frame color $B-K$\,$\sim$\,4.2, is lower than the very red colors ($B-K$\,$>$\,6) found among other known redshifted CO absorption systems at $z$\,$<$\,1. Nevertheless, when including also the many non-detections from the literature, a potential correlation between the absorption-line strength and $B-K$ color is evident, giving weight to the argument that the red colors of CO absorbers are due to a high dust content. 
\end{abstract}



\section{Introduction} 
\label{sec:intro}

Radio synchrotron jets emanating from active black holes are sources of bright continuum emission, which can act as background candles that facilitate the search for spectral lines in absorption at cm and mm wavelengths. As a result, radio absorption lines have provided a diagnostic for studing neutral and molecular gas clouds along our line-of-sight towards radio sources (see reviews by \citealt{com08} and \citealt{mor18}). 

Studies of molecular absorbers at low and intermediate redshifts provided insight into the raw materials that fuels early star formation and active galactic nuclear (AGN) activity \citep[e.g.,][]{ger97,wik99,wik18,all19,mac18,com19,ros19,ros23,mor23}. In addition, they have been used to study extra-galactic chemistry \citep[e.g.,][]{com97,mul11,mul14}, time-variability in the background radio source \citep[e.g.,][]{mul23}, the temperature of the Cosmic Microwave Background \citep[e.g.,][]{mul13,rie22}, and potential space-time variations of fundamental constants \citep[e.g.,][]{car00,cur11a,mul21}. When extended to higher redshifts, such studies could provide critical new insights into physical processes that govern cosmology and galaxy evolution in the Early Universe. Moreover, the strength of the absorption signal scales with the flux of the background radio continuum, and not directly with luminosity distance. This means that absorption lines have the potential to trace molecular clouds in the Early Universe down to gas masses that cannot be detected with emission-line observations, considering that even the most detailed high-$z$ emission-line studies only resolve molecular clumps containing millions to billions of solar masses of cold gas \citep[see, e.g.,][]{hod12,des23}.

To date, molecular hydrogen  absorption has been detected in over 30 damped Ly$\alpha$ systems at high redshifts \citep[e.g.,][]{lev85,bal14,not15,ran18,bal18}. However, in the millimeter regime, when it comes to lines-of-sight against discrete and typically strong background continuum sources, redshifted molecular absorption is rare: until now, CO had only been detected along six sight-lines at $z_{\rm abs}\ga0.1$, all of which were at $z_{\rm abs}\leq0.89$ \citep{wik95,wik96b,wik97,wik98,wik18,all19}.\footnote{Apart from this, the
highest redshift detection is absorption by H$_2$O at
$z_{\rm abs}=6.34$ against the CMB background with the Atacama
Large Millimeter/submillimeter Array (ALMA, \citealt{rie22}).} This
is despite extensive millimeter-band searches, both targeted
\citep{wik95,wik96a,dri96,mur03,cur08,cur11} and untargeted \citep{kan14,kli19}.\footnote{The work by \citet{kli19} is based on ALMA calibration data from \citet{ote16}.} A new study by \citet[][]{com23} reports the detection of three
new CO absorbers, two at $z_{\rm abs}\approx1.2$ and one at $z_{\rm abs}=3.387$.\footnote{\citet{com23} also report the detection of an HNC absorber at $z_{\rm abs}\approx1.3$.} The latter is an intervening absorber tentatively detected at $\sim3\sigma$, near the redshift of \HI\ absorption detected by \citet{kan07}.

Here we present the detection of molecular absorption associated with the high-redshift radio galaxy B2 0902+34 at $z$\,=\,3.396. B2 0902+34 was one of the first galaxies discovered in the Early Universe, thanks to its bright radio continuum, which served as a beacon for tracing the faint host galaxy \citep{lil88,eis92}. It was suggested that this is a proto-galaxy undergoing its first episode of star formation  \citep{eal93,pen99}. X-ray observations revealed that it contains a heavily obscured AGN \citep{fab02}. \citet{car95} described the radio/optical structure of this galaxy as ``bizarre'', with a bright northern radio lobe that shows a sharp ($\sim$90$^{\circ}$) bend and a flat-spectrum radio core located in a ``valley'' between two optical peaks. Recent 144 MHz imaging of the radio source with the Low Frequency Array (LOFAR) suggests that the complex radio structure is either the result of different episodes of AGN activity, or, more likely, due to interactions between the radio source and the surrounding halo gas \citep{cor23}. B2\,0902+34 is surrounded by a rich circumgalactic medium (CGM) in the form of a giant Ly$\alpha$ nebula \citep{reu03}, and is thought to be a collapsing protogiant elliptical \citep{ada09}. Searches for molecular gas using CO(4-3) emission were done only using single-dish telescopes, which set limits of M$_{\rm H_2}$\,$\la$\,5\,$\times$10$^{10}$ M$_{\odot}$ \citep{eva96,oji97}. B2\,0902+34 is exceptional in that it is one of the very few $z$\,$>$\,2 radio sources that has been observed to have an associated \HI\ absorption line (\citealt{cur08}; see also \citealt{uso91}, \citealt{bri93}, \citealt{cod03}, and \citealt{cha04}). Here we show that B2\,0902+34 also has molecular absorption lines corresponding to carbon monoxide, \CO, and the cyano radical, \CN. At $z$\,=\,3.396, it is the highest redshift CO absorber known to date.

Throughout this paper, we assume the following cosmological parameters: H$_{0}$\,=\,71 km\,s$^{-1}$\,Mpc$^{-1}$, $\Omega_{\rm M}$\,=\,0.27, and $\Omega_{\rm \lambda}$\,=\,0.73 \citep{wri06}. The corresponding angular scale is 7.3$^{\prime\prime}$ per arcsec.

\section{Data}
\label{sec:data}

The observations of B2\,0902+34 were performed with NSF's Karl G. Jansky Very Large Array (VLA) in D-configuration under project 21A-059 during 21 March $-$ 8 May 2021. The total on-source time was 18.7 hours. We used a continuous bandwidth coverage of 1\,GHz, consisting of 16 overlapping sub-bands of 128 MHz with 1\,MHz channels, centred around the redshifted \CO\ line at 26.2\,GHz ($\nu_{\rm rest}$\,=\,115.2712\,GHz). We observed the primary calibrator source J09027+3902 located at 6.5$^{\circ}$ distance from B2\,0902+34 every 5\,min to calibrate the complex gains and bandpass. 3C\,286 and 3C\,147 were used for absolute flux calibration.

The data were processed using the Common Astronomy Software Applications (CASA; \citealt{casa22}), using version 6.4.3-27 during the calibration and 6.5.2-26 for the imaging. After a standard manual calibration, we self-calibrated the data using the unresolved radio continuum source of B2\,0902+34 to further correct the complex gains. We then imaged the radio continuum using the line-free channels and applying a natural weighting scheme, which resulted in a synthesized beam of 3.9$^{\prime\prime}$\,$\times$\,3.5$^{\prime\prime}$ (PA -12.1$^{\circ}$). We also performed a deconvolution to clean the signal of the radio continuum. The radio continuum is unresolved and peaks at 22.1\,$\pm$\,1.1 mJy\,beam$^{-1}$.

For the line data, artifacts appeared at the level of a few per cent, which is smaller than the assumed 5$\%$ uncertainty in the flux calibration of the VLA data. These artifacts are likely related to small variations in the point-spread function (PSF) across the large fractional bandwidth of the data or inaccuracies in the relative calibration between the 16 sub-bands, combined with the limited spectral dynamic range reached by our observations (1 in a few 100). These artifacts scale with the brightness of the continuum emission. As a result, they become noticeable against the peak of the radio continuum as small (mJy-level) amplitude `jumps' along the bandpass. To mitigate this effect to a level that it does not affect the line data, we created a continuum-free ($u$,$v$)-data set prior to imaging the line data. For this, we used the continuum model that was derived during the deconvolution process of the continuum imaging, which we converted into model visibilities by applying a Fourier transform with the CASA task `ft'. These model visibilities were then subtracted from the ($u$,$v$)-data with the CASA task `uvsub'. Any residual continuum emission was subsequently subtracted in the ($u$,$v$)-domain using a linear fit to the line-free channels with CASA task `uvcontsub'. Our \CO\ absorption was well captured in a single sub-band, in a region of the spectrum that did not suffer from any of the artifacts that appeared prior to the continuum subtraction. This allowed us to verify that our continuum subtraction did not negatively affect the absorption signal (see Appendix \ref{sec:contsub}).

After subtracting the continuum, we imaged the line data using a natural weighting scheme and native 11.9 \kms\ channels, while cleaning the line signal in each channel until an absolute value for the threshold of 0.35 mJy\,beam$^{-1}$ was reached. The root-mean-square (rms) noise of this image cube is $\sigma$\,=\,0.05 mJy\,beam$^{-1}$\,channel$^{-1}$.

\subsection{HST}

We also obtained an image with the {\it Hubble Space Telescope (HST)} Wide Field Camera 3 (WFC3) in the F105W filter (project ID: 17268, PI Emonts). The observations were executed on 9 October 2023, and the total on-source integration time was 37\,min. The F105W filter is devoid of emission lines at the redshift of B2\,0902+34 and traces the stellar continuum at 300\,nm in the rest-frame. We obtained the F105W image from the Multimission Archive at the Space Telescope Science Institute (MAST): \dataset[10.17909/cy88-7e89]{https://doi.org/10.17909/cy88-7e89}. A detailed analysis of the {\it HST} imaging will be postponed to a future paper.

\section{Results}
\label{sec:results}

\begin{figure*}
\centering
	\includegraphics[width=\textwidth]{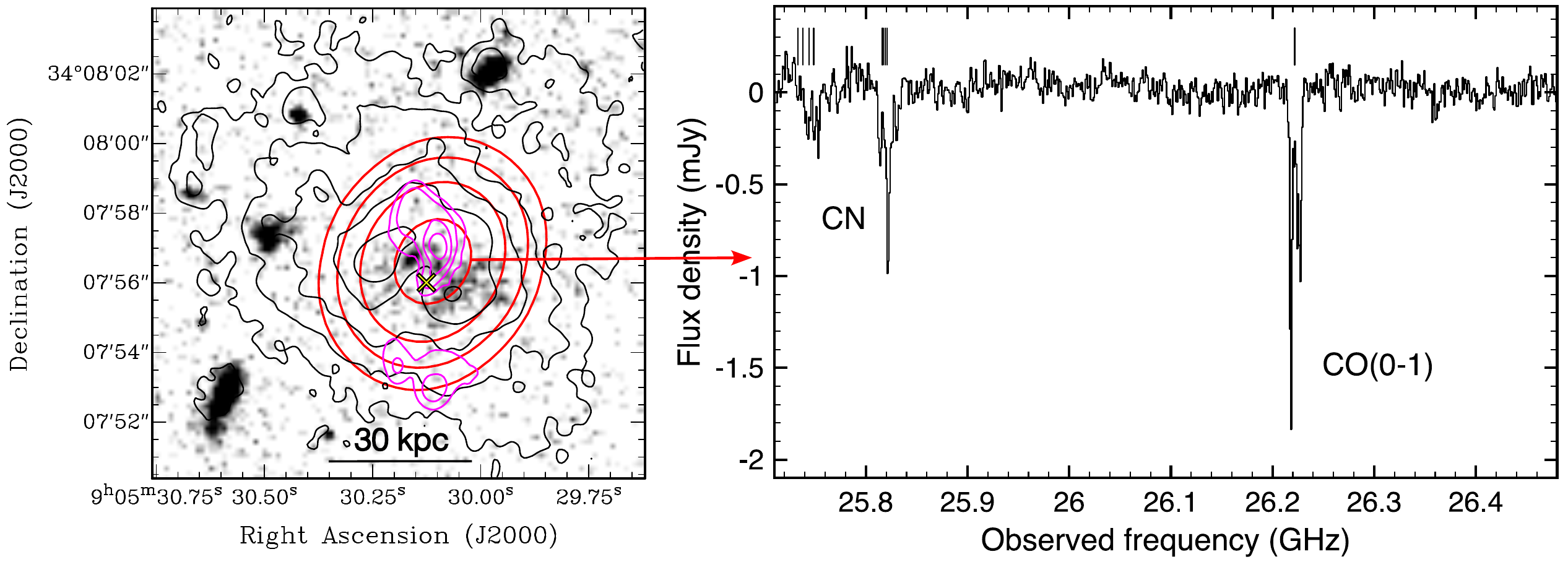}
    \caption{VLA observations of B2\,0902+34. Left: {\it HST} WFC3/F105W image of B2\,0902+34, with overlaid in red the contours of the unresolved continuum emission of our VLA D-configuration data, and in magenta contours of the 144\,MHz Low Frequency Array (LOFAR) image from \citet{cor23}. The contour levels of the VLA [LOFAR] data start at 2.2 [2.5] mJy\,beam$^{-1}$ and increase by a factor 2 [4]. The yellow cross marks the location of the nucleus identified by \citet{car95}. We shifted the LOFAR data by $\sim$0.7$^{\prime\prime}$ to match this astrometry. The black contours represent the Ly$\alpha$ image from \citet{reu03}, with levels starting at 4$\%$ of the peak flux in the halo and increasing by a factor of 2. Right: Spectrum of B2\,0902+34 taken against the unresolved VLA continuum. The flux density after continuum subtraction is plotted against the observed frequency. The small vertical bars above the spectrum indicate the redshift $z$\,=\,3.3960 for the \CO\ and two \CN\ lines, which in turn consist of a series of hyperfine lines.}
    \label{fig:spectrum}
\end{figure*}

Figure \ref{fig:spectrum} shows the absorption spectrum taken against the radio continuum of B2\,0902+34, where the lines of \CO, \CN\ $J$=3/2-1/2, and \CN\ $J$=1/2-1/2 are clearly evident. While the background continuum source has a total extent of $\sim$5$^{\prime\prime}$ \citep{car95}, the nucleus and brightest part of the northern lobe are unresolved in our D-configuration observations, while the faint southern radio lobe is not detected in our 26\,GHz data. The peak flux-density of this unresolved background continuum is 22.1\,$\pm$\,1.1 mJy\,beam$^{-1}$.\footnote{The uncertainty is based on an assumed 5$\%$ uncertainty in the absolute flux calibration.}

The {\it HST} WFC3/F105W image in Fig.\,\ref{fig:spectrum} (left) shows faint stellar light across a region of $\sim$30 kpc, similar to the low surface-brightness emission seen by \citet{pen99} in a bluer F622W ($\lambda_{\rm rest}$\,$\sim$\,180\,nm) image taken with the Wide Field and Planetary Camera 2 (WFPC2). This optical emission is faintest at the location of the radio core, which could hint to the presence of large amounts of dust \citep[see also][]{pen99}. However, the astrometry remains somewhat ambiguous, because all the radio maps relied on self-calibration, which leaves inherent astrometric errors \citep[e.g.,][]{pea84}. A detailed discussion of the {\it HST} data will be given in a future paper.

\subsection{\CO\ absorption}
\label{sec:CO01}

\begin{figure}
\centering
	\includegraphics[width=\columnwidth]{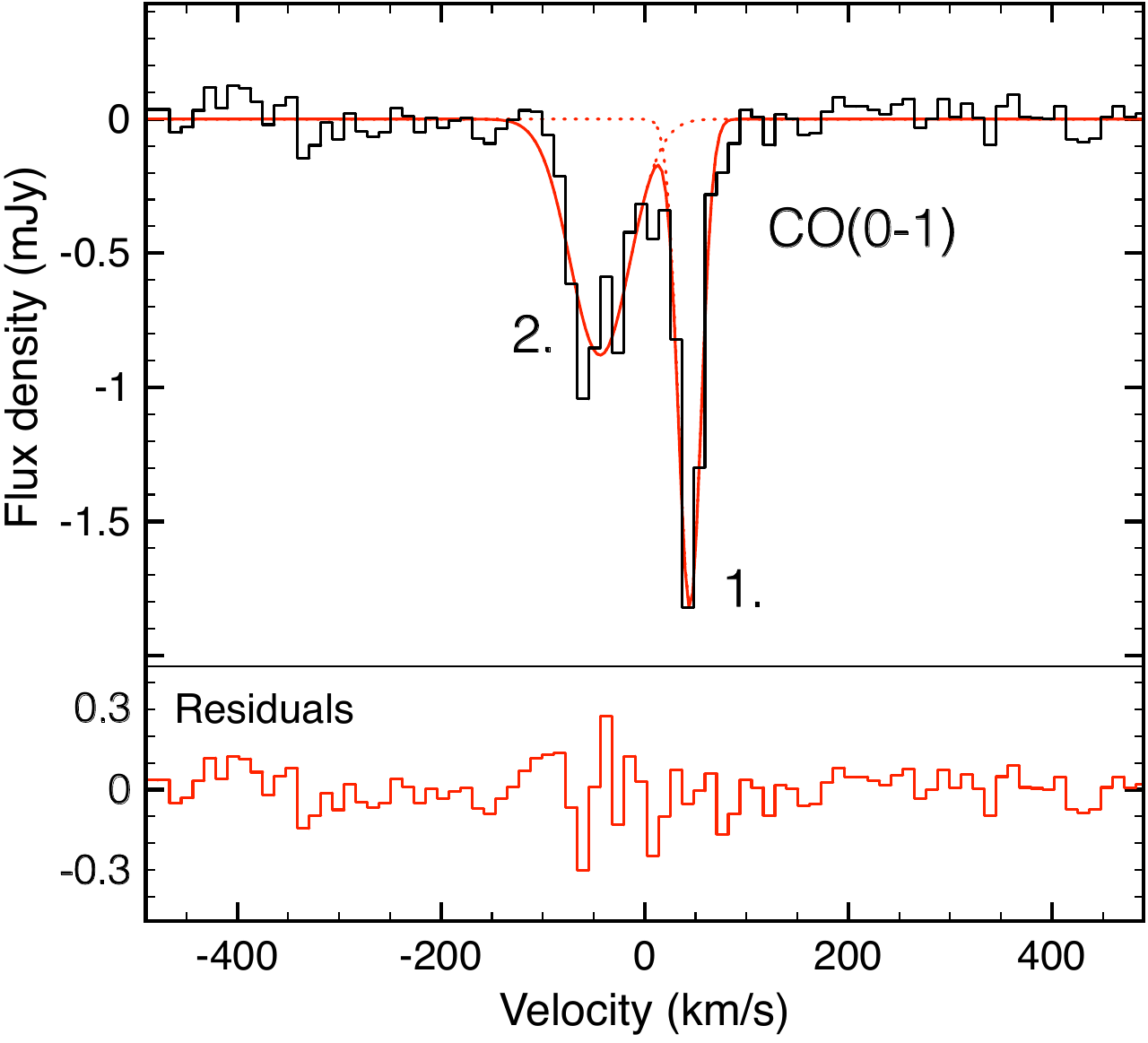}
    \caption{Spectrum of \CO. The flux density after continuum subtraction is plotted against the radio velocity with respect to $z$\,=\,3.3960. The red line shows a double Gaussian fit, with the dashed lines the individual Gaussian components (indicated with 1 and 2, as per Table \ref{tab:results}). The bottom panel shows the residuals after subtracting the fitted model from the spectrum.}
    \label{fig:CO10}
\end{figure}

Figure \ref{fig:CO10} shows in detail the \CO\ absorption, which consists of two components (Table\,\ref{tab:results}). We place the systemic redshift in the middle between the two components, namely $z$\,=\,3.3960\,$\pm$0.0008. The uncertainty reflects the difference in redshift of the two CO components. 

The deep, narrow CO component has a redshift of $z_{\rm CO}$\,=\,3.3966\,$\pm$0.0001, which is in agreement with the redshift of $z_{\rm HI}$\,=\,3.3967\,$\pm$0.0002 derived from \HI\ 21-cm absorption of neutral hydrogen gas that was previously detected with the Giant Meterwave Radio Telescope (GMRT; \citealt{cha04}), as well as earlier detections with the VLA \citep{uso91}, the Arecibo telescope \citep{bri93}, and the Westerbork Synthesis Radio Telescope (WSRT; \citealt{cod03}). This CO component has a FWHM$_{\rm CO}$\,$\approx$\,26 \kms, which is a factor three narrower than the FWHM of the \HI\ absorber \citep{cha04}. The optical depth of the narrow CO component is $\tau$\,$\ga$\,8.6$\%$. Our observed optical depth is based on the unresolved background continuum at 26 GHz. Unless the absorption covers the background radio-continuum source uniformly, resolving the radio continuum at higher spatial resolution would decrease the flux density of the background continuum against which the absorption occurs. This is why our estimate of the observed optical depth $\tau_{\rm obs}$ is a lower limit to the true optical depth $\tau$. 

\begin{figure}
\centering
	\includegraphics[width=\columnwidth]{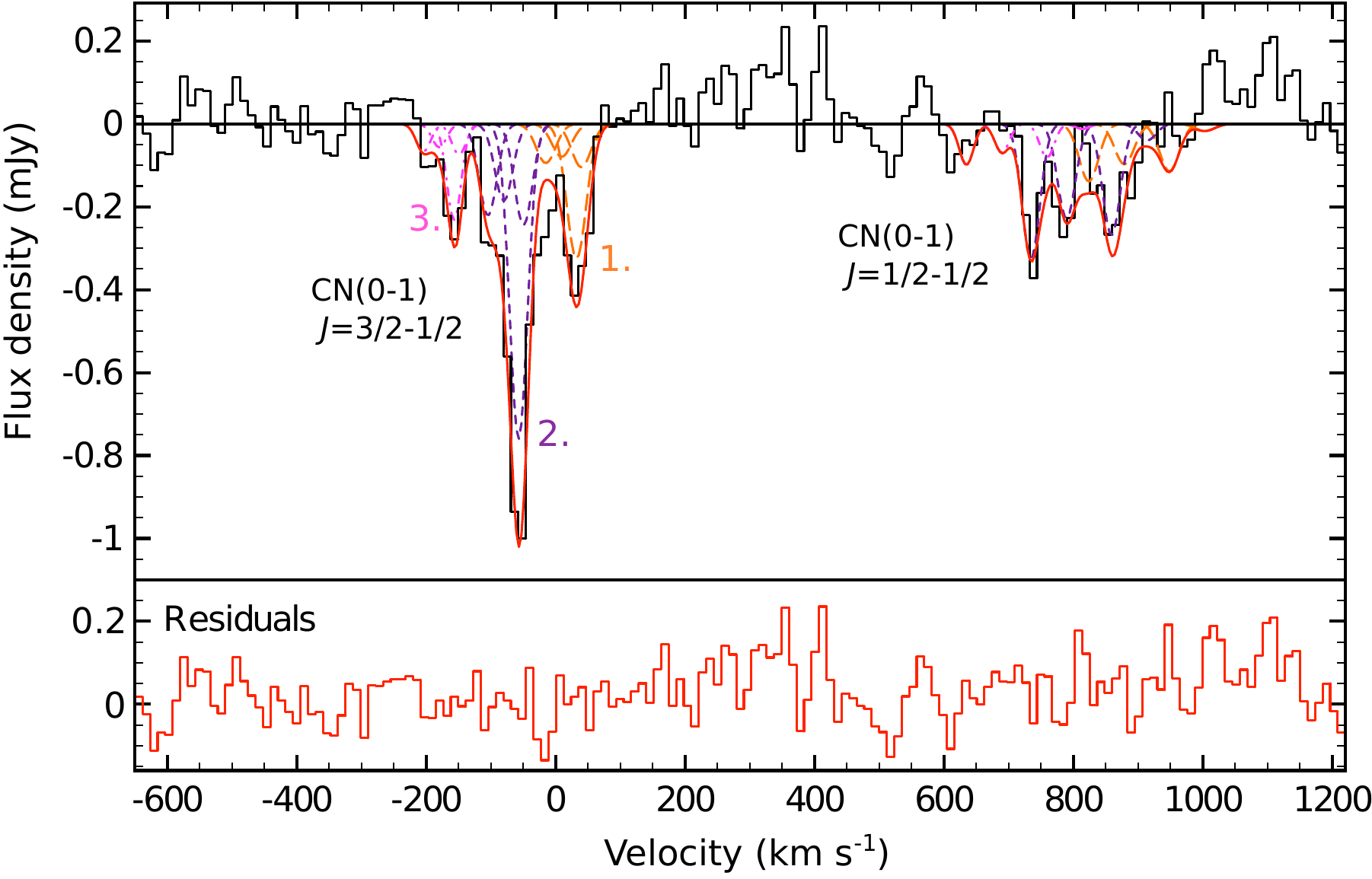}
    \caption{Spectrum of \CN. The flux density after continuum subtraction is plotted against the radio velocity of the stronger \CN\ $J$=3/2-1/2 multiplet with respect to $z$\,=\,3.3960. The red line shows a fitted model consisting of three components, with each component representing a CN absorber (orange, purple, and magenta colors represent components 1, 2, and 3 from Table \ref{tab:results}, respectively). Each component in turn consists of four Gaussians (visualized by the same color), which represent the four detectable hyperfine lines of each CN absorber. The bottom panel shows the residuals after subtracting the fitted model from the spectrum.
    }
    \label{fig:CN}
\end{figure}

A broader \CO\ component is centered on a velocity of -88 \kms\ blueward of the narrow line, with FWHM$_{\rm CO}$\,$\approx$\,69 \kms\ and $\tau$\,$\ga\,4.1$\%. This CO component has no obvious counterpart in the \HI\ 21-cm absorption spectra.

\subsection{\CN\ absorption}
\label{sec:CN}

\CN\ consists of nine hyperfine lines in the VLA band, divided into two groups: the \CN\ $J$=3/2-1/2 group has the highest integrated brightness and consists of five lines, one of which is too faint to be detected above the noise in our data. The fainter \CN\ $J$=1/2-1/2 group consists of four lines. As shown in Fig.\,\ref{fig:CN}, three components, representing three different absorbers, are required to obtain a good fit to the CN spectra. The fitting was performed with an unconstrained Gaussian fit to the brightest (hereafter ``main'') hyperfine line for each of the three components of the stronger \CN\ $J$=3/2-1/2 multiplet, while at the same time representing the other hyperfine lines with additional Gaussians that had their center, width, and intensity constrained to the main line as per atomic physics \citep{ost89}. The exact same solutions were applied to the hyperfine lines of the weaker \CN\ $J$=1/2-1/2 multiplet, again constraining the center, width, and intensity based on atomic physics. Therefore, the fully constrained fit to the weaker \CN\ $J$=1/2-1/2 multiplet merely serves to assure that our model accurately represents both CN multiplets (see Fig.\,\ref{fig:CN} for details). 

\begin{figure}
\centering
	\includegraphics[width=\columnwidth]{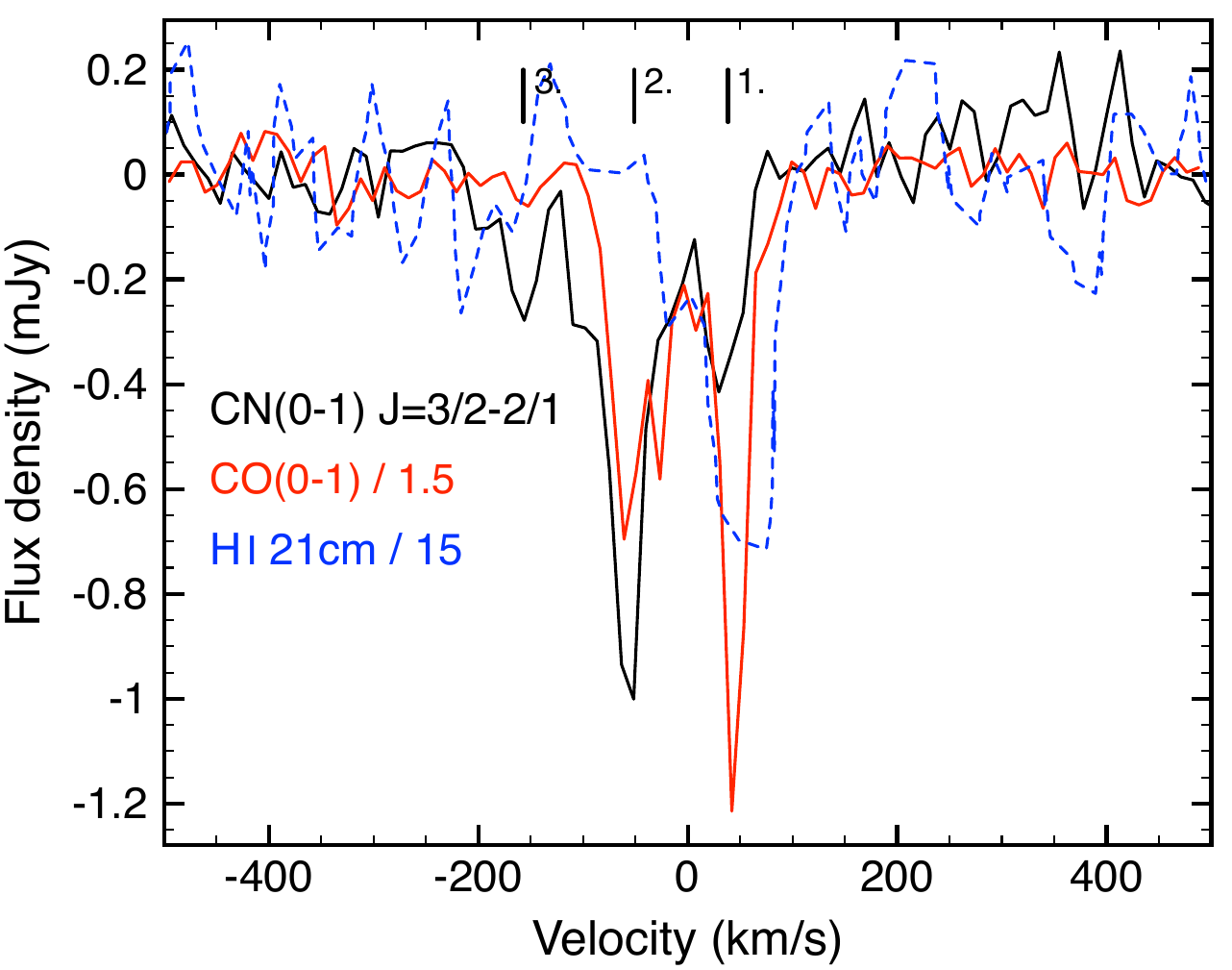}
    \caption{Overlay of \CN\ $J$=3/2-1/2 (black line), \CO\ (red line), and the \HI\ 21cm spectrum from \citet{cha04} (blue dashed line). For clarity, the CO(1-0) and \HI\ spectra were scaled down by a factor of 1.5 and 15, respectively. The three components from Table \ref{tab:results} are indicated with the small vertical bars.}
    \label{fig:overlay}
\end{figure}

The redshifts of components 1 and 2 are consistent to within one channel with the redshifts of the two \CO\ absorbers (Fig. \ref{fig:overlay}). We therefore assume that the CN and CO absorptions originate from the same gas reservoir.  Component 3  of CN is the weakest of the three and has no counterpart in \CO\ at the detection limit of our data.

\begin{deluxetable*}{lcccccccc}
\tablecaption{Results of Gaussian fitting to the spectra.}
\tablehead{
\colhead{Line} & \colhead{Comp.} & \colhead{$z$} & \colhead{$v^{*}$} & \colhead{$FWHM^{\dagger}$} & \colhead{$S_{\rm abs}$} & \colhead{$\tau_{\rm obs}^{\ddagger}$} & \colhead{$\int\tau_{\rm obs} \delta v^{\S}$} & \colhead{CO/CN$^{\P}$} \\
\colhead{} & \colhead{} & \colhead{} & \colhead{(\kms)} & \colhead{(\kms)} & \colhead{(mJy)} & \colhead{} & \colhead{(\kms)} & \colhead{}
}
\startdata
\CO\ & 1 & 3.3966$\pm$0.0001 & 44$\pm$6 & 26$\pm$6 & -1.82$\pm$0.12 & 0.086$\pm$0.007 & 2.98$\pm$0.73 & \\
 & 2 &  3.3953$\pm$0.0001 & -44$\pm$6 & 69$\pm$8 & -0.88$\pm$0.07 & 0.041$\pm$0.004 & 2.36$\pm$0.36 & \\
  & & & & & & & \\ 
\CN\ & 1 & 3.3965$\pm$0.0001 &  32$\pm$7  & 35$\pm$10 & -0.33$\pm$0.07 & 0.015$\pm$0.003 & 1.24$\pm$0.45 & 2.4 \\
 & 2 & 3.3952$\pm$0.0001 & -57$\pm$6  & 32$\pm$7 & -0.76$\pm$0.08 & 0.035$\pm$0.004 & 2.32$\pm$0.58 & 1.0 \\
& 3 & 3.3938$\pm$0.0001 & -157$\pm$10 & 26$\pm$10 & -0.23$\pm$0.07 & 0.010$\pm$0.003 & 0.66$\pm$0.33 & $\le$0.4\\
\enddata
\tablecomments{
  Uncertainties include uncertainties in the fitting, as well as half the width of a channel (for $v$ and $FWHM$) and a 5$\%$ uncertainty in absolute flux calibration (for $S_{\rm abs}$). Uncertainties are added in quadrature, and propagated for $\tau_{\rm obs}$ and $\int\tau_{\rm obs} \delta v$.\\
$^{*}$ Velocity is with respect to $z$\,=\,3.3960 (v\,=\,0 \kms).\\
$^{\dagger}$ $FWHM$ is the full width at half the maximum intensity.\\
$^{\ddagger}$ The observed optical depth, $\tau_{\rm obs}$, is estimated using $\tau_{\rm obs}$\,=\,-ln(\( \frac{S_{\rm cont}-S_{\rm abs}}{S_{\rm cont}} \)). Because the background continuum is unresolved, $\tau_{\rm obs}$ represents the lower limit to the true optical depth, $\tau$.\\
$\S$ The observed integrated optical depth, $\int\tau_{\rm obs} \delta v$, for CN reflects the combined value of all the hyperfine lines of the \CN\ $J$\,=\,3/2-1/2 group. Following atomic physics, the integrated optical depth of the model fit to the 1/2-1/2 group was constrained to be a factor 0.61 lower than that of the 3/2-1/2 group. The contribution of the 1/2-1/2 group is not included in the above estimate of $\int\tau_{\rm obs} \delta v$, to facility easy comparison with results from the literature.\\
$\P$ The CO/CN line ratio is defined as the ratio of the integrated optical depths, $\int\tau_{\rm CO,\,obs}\,\delta v_{\rm CO}$/$\int\tau_{\rm CN,\,obs}\,\delta v_{\rm CN}$.
}
\label{tab:results}
\end{deluxetable*}

\section{Discussion}
\label{sec:discussion}

\subsection{Physical properties of the molecular gas}
\label{sec:properties}

The detection of both \CO\ and \CN\ in absorption allows us to derive physical properties of the absorbing gas. The ground-transition \CO\ is the most reliable tracer for the overall molecular gas content across the full range of densities, independent of the excitation properties of the gas \citep[e.g.,][]{bol13}. Contrary, \CN\ traces moderately dense ($\ga$10$^{4}$ cm$^{-3}$) molecular gas \citep{bro14,shi15}.

\subsubsection{Column densities}
\label{sec:columndensity}

To derive CO column densities, we follow \citet[]{all19}:\footnote{See also \citet{wil13}, \citet{wik95}, \citet{man15}, and \citet{ros19}}
\begin{equation}
N_{\rm CO} \ge \frac{8\pi}{c^3} \frac{{\nu}^{3}}{g_{J+1} A_{J+1}} \frac{Q(T_{\rm ex})e^{E_{J}/k_{B}T_{\rm ex}}}{1-e^{-h\nu/k_{B}T_{\rm ex}}} \int \tau_{\rm CO} \delta v ,
\end{equation}
where $\nu$ is the \CO\ rest frequency, $g_{J+1}$ the statistical weight for the upper ($J+1$\,=\,1) energy level, $A_{J+1}$\,=\,7.67$\times$10$^{8}$ s$^{-1}$ the Einstein A-coefficient  \citep[see][]{cha96}, $Q({T_{\rm ex})}$ the partition function assuming a single excitation temperature $T_{\rm ex}$, $E_{J}$ the energy of the lower ($J$\,=\,0) level, and $\int\tau_{\rm CO}\delta v$ the optical depth integrated over the width of the absorption profile (from Table \ref{tab:results}). The variables $h$ and $k_{B}$ are the Planck and Boltzmann constants, respectively. For CO, the rigid-rotor approximation allows us to use $g_{J+1}$\,=\,2$(J+1)$+1, and also approximate $Q$($T_{\rm ex}$)\,$\approx$\,$T_{\rm ex}$/B (for $T_{\rm ex}>>B$) and $E_{J}$/$k_{b}$\,$\approx$\,$J$($J+1$)$B$ \citep[see][]{all19}. Here, $B$\,=2.766\,K is the rotational constant for CO. Furthermore, we assume $T_{\rm ex}$\,$\ga$\,15\,K \citep[e.g.,][]{wil97}. Higher values than the lower limit of 15\,K are expected if the molecular gas is located in star-forming regions or the vicinity of the AGN. The above parameters imply CO column densities of $N_{\rm CO}$\,$\ge$\,2.4\,$\times$10$^{16}$ cm$^{-2}$ for the deep, narrow component, and $N_{\rm CO}$\,$\ge$\,3.1\,$\times$10$^{16}$ cm$^{-2}$ for the broader CO component. We do not take into consideration potential line-dimming as a result of the increased temperature of the Cosmic Microwave Background radiation at $z$\,=\,3.4 \citep{dac13,zha16}. This provides an additional reason for considering $N_{\rm CO}$ to be a lower limit.

If we assume a CO/H$_{2}$ abundance of 1\,$\times$\,10$^{-4}$ that was found for molecular clouds in our Milky Way Galaxy \citep{fre82}, then each component has a total H$_{2}$ column density of roughly $N_{\rm H_2}$\,$\ga$\,3$\times$10$^{20}$ cm$^{-2}$. 

\subsubsection{CN abundance}
\label{sec:CNabundance}

The formation of CN occurs through several routes that take place simultaneously in molecular clouds, especially at the transition boundary from \HI\ to H$_{2}$ in regions illuminated by ultra-violet (UV) radiation. These routes include photo-dissociation of HCN (HCN + $\nu$ $\rightarrow$ CN + H), collisions with hydrogen atoms in high-density regions (HCN + H $\rightarrow$ CN + H$_{2}$), dissociative recombination (HCN$^{+}$ + e$^{-}$ $\rightarrow$ CN + H), and chemistry in photo-dissociation regions (PDRs) at an extinction of A$_{\rm V}$\,$\sim$\,2 mag (N + C$_{2}$ $\rightarrow$ CN + C, and N + CH $\rightarrow$ CN + H) \citep[e.g.,][]{ste95}. This means that the CN abundance is enhanced in the outer regions of molecular clouds exposed to UV radiation fields in PDRs \citep[e.g.,][]{fue95,aal02,bog05}, or in regions where the X-ray ionization rates are high \citep[e.g.,][]{mei07,lep96}.

\begin{figure}
\centering \includegraphics[width=\columnwidth]{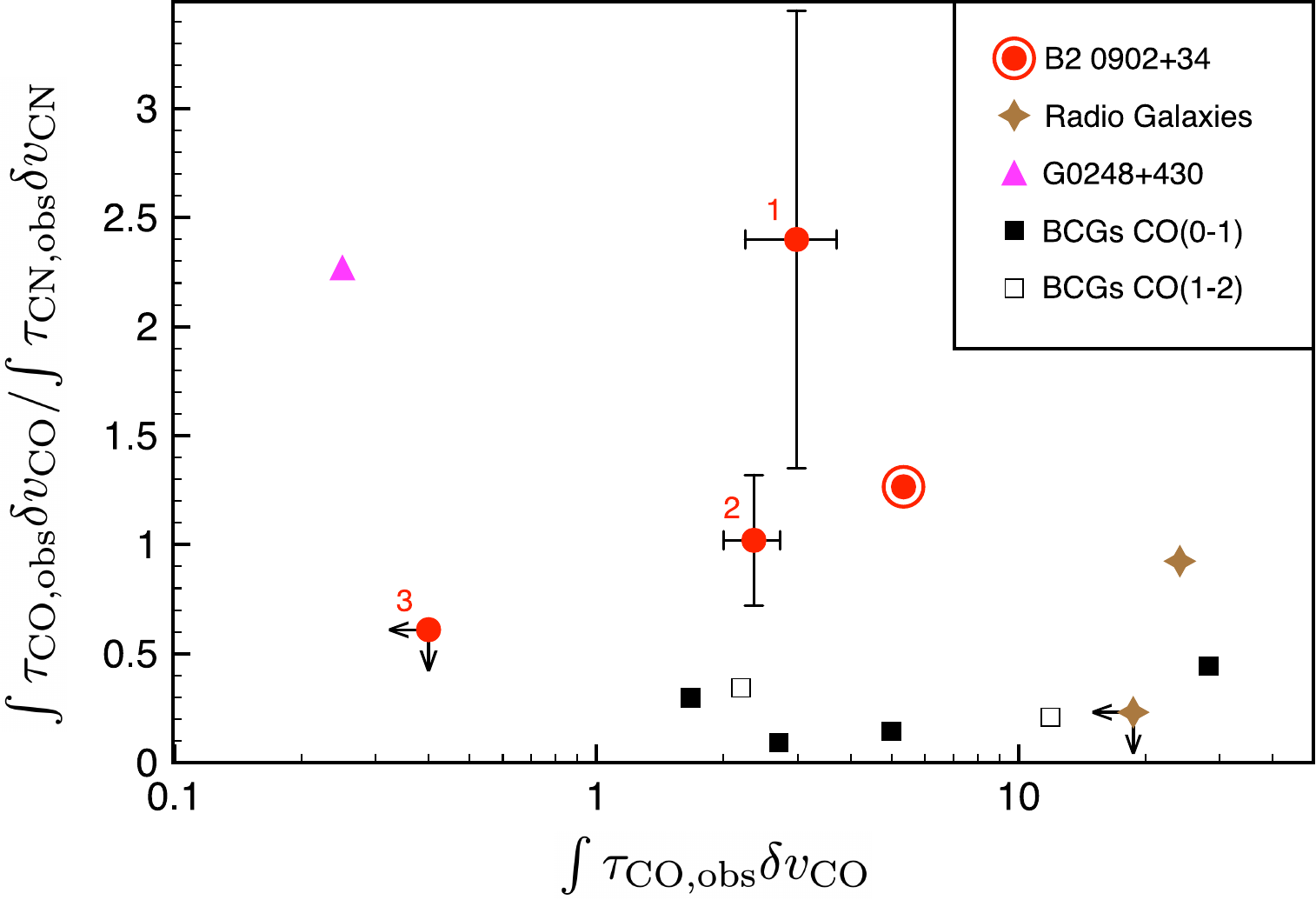}
\caption{Ratio of the observed integrated optical depths ($\int\tau_{\rm obs} \delta v$) of \CO\ and \CN\ $J$\,=\,3/2-1/2, plotted against the integrated optical depth of \CO. The circled red dot represents the average value for B2\,0902+34, with the red dots the three individual components from Table \ref{tab:results}. The other symbols represent low-$z$ systems: the brown stars are two radio galaxies, namely 4C\,31.04 at $z$\,=\,0.06 and upper limits for 4C\,52.37 at $z$\,=\,0.1 \citep[][also Murthy et al. in prep.]{mor23}; the magenta triangle is the merger system G0248+430 at $z$\,=\,0.05 \citep{com19}; squares represent Brightest Cluster Galaxies (BCGs) at 0.01\,$\la$\,$z$\,$\la$\,0.23, with the two open squares BCGs observed in CO($J$=1$\rightarrow$2) instead of \CO\ \citep{ros19}.}
\label{fig:CO_CN}
\end{figure}

The observed ratios of the velocity-integrated optical depths of the \CO\ and \CN\ $J$\,=3/2-1/2 lines, or $\int\tau_{\rm CO, obs} \delta v_{\rm CO}$/$\int\tau_{\rm CN, obs} \delta v_{\rm CN}$ (hereafter CO/CN), are CO/CN\,$\sim$\,2.4 for component 1, CO/CN\,$\sim$\,1.0 for component 2, and CO/CN\,$\la$\,0.4 for component 3, with the latter assuming a \CO\ non-detection of 3$\sigma$ across FWHM$_{\rm CO}$\,$\equiv$\,FWHM$_{\rm CN}$\,=\,26 \kms\ (see Table \ref{tab:results}). As can be seen in Fig.\,\ref{fig:CO_CN}, the CO/CN value for component 1 is comparable to the absorption-line value from a tidal tail in merger system G0248+430, which is illuminated by a background quasar \citep{com19}. The CO/CN value for component 2 is similar to the value found in the low-$z$ radio galaxy 4C\,31.04 \citep[][see also Murthy et al. in prep.]{mor23}. Component 3 has a CO/CN limit comparable to the limit found for radio galaxy 4C\,52.37 \citep{mor23}. Given that B2\,0902+34 is thought to be a collapsing protogiant elliptical \citep{ada09}, we also compare the CO/CN absorption values with those of a small sample of low-$z$ Brightest Cluster Galaxies (BCGs) studied by \citet{ros19}. These low-$z$ BCGs have CO/CN ratios in agreement with the limit that we derive for component 3 in B2\,0902+34, but lower than those of the main components 1 and 2.

When we compare our absorption results to emission-line studies, our CO/CN values are low compared to emission-line ratios found in nearby galaxies \citep{wil18} and luminous infra-red galaxies \citep[LIRGs;][]{aal02}, which typically have CO/CN\,$\ga$\,10. However, our absorption results are comparable to emission-line ratios found in the molecular outflow of Mrk 231 \citep{cic20}, the circum-nuclear disk of the active galaxy NGC\,1068 \citep{gar10}, and the $z$\,$\sim$2.56 Cloverleaf quasar \citep{rie07}. This suggests that the CN abundance is likely enhanced by a UV or X-ray radiation field, and may thus originate close to the central AGN region, which is designated as region `N' by \citet{car95}. In Mrk 231, the outflow component shows roughly 4$\times$ lower CO/CN ratios than gas in the center of the host galaxy \citep{cic20}. For absorption detections, any outflow would be aligned in front of the radio source, and thus be blueshifted. In this respect, it is interesting that the most blueshifted component in B2\,0902+34 also has the lowest CO/CN value. However, regarding all the above, we note that a one-on-one comparison between emission- and absorption-line studies may be limited by density and beam-filling effects. Moreover, the CN abundance is very sensitive to the fraction of mechanical energy in photon-dominated regions \citep{kaz15}.

Alternatively, \citet{mor23} discuss that CN can be enhanced relative to CO in low-density, irradiated molecular shocks \citep{god19,leh22}, or diffuse molecular gas with a high velocity dispersion \citep{wak15}. The CO/CN values that we find in B2\,0902+34 resemble those found in the low-$z$ radio galaxies 4C\,31.04 and 4C\,52.37 by \citet{mor23} (Fig.\,\ref{fig:CO_CN}). This could be consistent with the scenario that the radio source interacts with the surrounding gaseous medium \citep[][see Sect.\,\ref{sec:intro}]{cor23}. 

In future work we will further address the nature of the CN absorption, by studying the location of the absorption components, and targeting other dense molecular gas tracers, such as HCN and HCO$^{+}$.

\subsubsection{Cloudy gas reservoir}
\label{sec:cloud}

The properties of the molecular gas reservoirs that cause the absorption features along the line-of-sight towards the radio continuum depend on whether the gas is part of the inter-stellar medium (ISM) close to the AGN, the ISM further out in the host galaxy, or the large-scale CGM. We will address this with future observations at higher spatial resolution, which will resolve the background continuum and therefore identify the location of the absorption components. For the moment, we will follow the assumption made in Sect.\,\ref{sec:CNabundance} that the molecular gas is part of the ISM, possibly near the AGN. We also follow \citet{gus20}, who show that the properties of molecular clouds (mass, velocity dispersion, and volume density) do not significantly change from redshift 4 to 0. The discussion in this Section will focus on the two strongest absorption components that are detected in \CO.

The estimated H$_{2}$ column densities traced by these two strong absorption components are $N_{\rm H_2}$\,$\ga$\,3$\times$10$^{20}$ cm$^{-2}$ (Section \ref{sec:columndensity}). Average volume densities of molecular clouds have been found to vary from roughly 1 - 1000 cm$^{-3}$ \citep[e.g.,][]{smi00,hei08,ben13,sch22}, with densities reaching $\sim$10$^{4-5}$ cm$^{-3}$ at the threshold for star formation \citep[e.g.,][]{lad10,ben13,baa23}. If a single molecular cloud along the line-of-sight towards the background radio continuum causes an absorption component, and we assume a traditional estimate of the volume density of $n_{\rm H_2}$\,$\sim$\,150 cm$^{-3}$ \citep{sco75}, then the diameter of this cloud would be $D$\,$\sim$\,0.7\,pc, with a lower limit to the mass of M$_{\rm H_2}$\,$\ga$\,1\,$M_{\odot}$. It is likely that local volume densities are higher and the cloud diameter smaller, because we detect the absorption components also in CN, which has an effective density of order $\ga$10$^{4}$ cm$^{-2}$ \citep{shi15}. This is a very simplistic view, given that molecular clouds are complex systems, with often highly structured, filamentary, or even fractal characteristics \citep[e.g.,][]{el96,com98,gol08,men10,hey15,sch20,won22,fah23,li23}. Nevertheless, it serves to illustrate that the absorption in B2\,0902+34 likely originates from small ($\la$pc) clouds rather than giant ($\sim$10-100\,pc) molecular cloud complexes. 

A single cloud likely has a velocity dispersion ($\sigma$) of order a few \kms\ \citep{sol87,ben13,miv17,spi22}. This is significantly smaller than the FWHM\,$\sim$\,26$-$69 \kms\ ($\sigma$ \,$\sim$\,11$-$29 \kms) of the absorption components that we detect in B2\,0902+34 (Table \ref{tab:results}). This suggests that the absorption components may be caused by regions with many small clouds or cloud fragments (`cloudlets'; \citealt[][]{git03}), where the overall kinematics are dominated by the velocity dispersion between the clouds. It is likely that the molecular clouds or cloudlets are optically thick, in which case their covering fraction of the background continuum is much less than 1. The crossing time of a cloud with a velocity of a few tens of \kms\ and a size $\la$1\,pc is short, $t_{\rm cross}$\,$\la$\,10$^{5}$ years. This means that the clouds are likely short-lived.

The absorption components that we detect in B2 0902+34 differ in CN abundance compared to CO (Fig.\,\ref{fig:CO_CN}). This could mean that the molecular gas structures are not uniform across the regions, and that perhaps high-density clouds or cloudlets are dispersed and mixed with lower density molecular gas. This could represent a ``cloudy region'', with many small and short-lived clumps covering a region that is highly turbulent. In such a region, cooling and dissipation would allow for the formation and growth of molecular clouds \citep[e.g.,][]{li14a,kanj21}, or the shattering of dense cold clouds could create a reservoir of diffuse and warmer molecular gas \citep[e.g.,][]{app23}. 

The fact that absorption component $\#$1 has approximately the same redshift as the previously detected \HI\ absorber (Fig.\,\ref{fig:overlay}) suggests that the neutral gas likely also originates from the same region. However, the factor three difference in line width between the deep CO absorption and the \HI\ absorption (Fig.\,\ref{fig:overlay}) suggests that the cool neutral gas is even more turbulent than the ensemble of cold gas clouds. Again, this would suggest a region where dense, molecular gas clouds are embedded in a larger reservoir of more diffuse gas.

In the presence of a powerful radio source, such a region could have properties as predicted by precipitation models of many small cloudlets distributed throughout a larger reservoir of gas, where the formation and destruction of molecular clouds is regulated through feedback \citep[e.g.,][]{sha10,sha12,li14b,voi15}.

\subsection{Comparison to previous CO absorption studies}

\begin{figure}
\centering \includegraphics[width=\columnwidth]{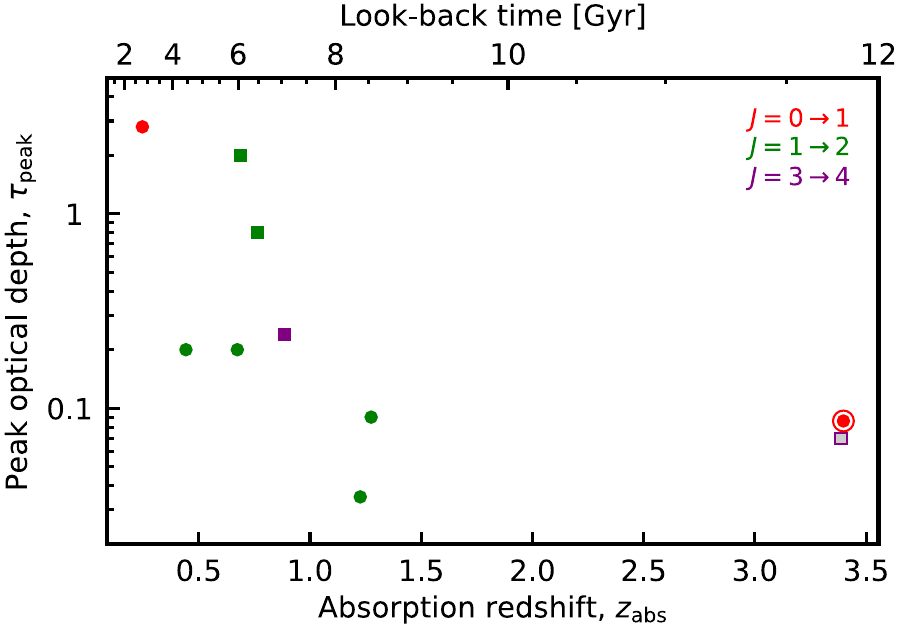}
\caption{The observed peak optical depth of the CO absorption versus the redshift for all the known redshifted CO absorbers (Table \ref{tab:mags}). Dots and squares represent the associated and intervening absorbers, respectively. The circled red dot represents B2\,0902+34. The purple-gray square is the tentative CO($J$=3$\rightarrow$4) absorption towards 0201+113 \citep{com23}.}
\label{fig:tau_z}
\end{figure}

Table \ref{tab:mags} gives an overview of the ten CO absorbers detected at $z$\,$>$\,0.1 (see Sect. \ref{sec:intro}). As shown in Fig. \ref{fig:tau_z}, the CO absorption in 0902+34 is the highest redshift example yet detected, possibly by a large margin. It is comparable in redshift only to the tentative detection of intervening CO($J$=3$\rightarrow$4) absorption towards 0201+113, and its look-back time of almost 12 Gyr is at least 3 Gyr earlier than the other known CO absorbers.

As discussed in Sect.\,\ref{sec:intro}, there has been a lack of CO absorbers against bright radio-continuum sources at high and intermediate redshifts, despite dedicated  searches. The high redshift of B2\,0902+34 allows detection by the VLA, which has a high sensitivity in comparison with the mm-band telescopes used previously. However, the integrated signal of the CO absorption in B2\,0902+34 is detected at a level of $\sim$45$\sigma$ after 18.7h of on-source integration time. This means that the absorption signal would be detectable with the VLA at a 5$\sigma$ level after only $\sim$14\,min on-source exposure time, which is similar to the sensitivity of previous searches for CO absorption in radio sources at $z$\,$>$\,1 \citep[e.g.,][]{cur11}. We therefore do not expect that the high sensitivity of our VLA observations introduced a bias or prevents a fair comparison with previous CO absorption studies in terms of detectability of the signal.

\subsubsection{Reddening versus absorption strength}
\label{sec:reddening}

\begin{figure}
\centering \includegraphics[width=\columnwidth]{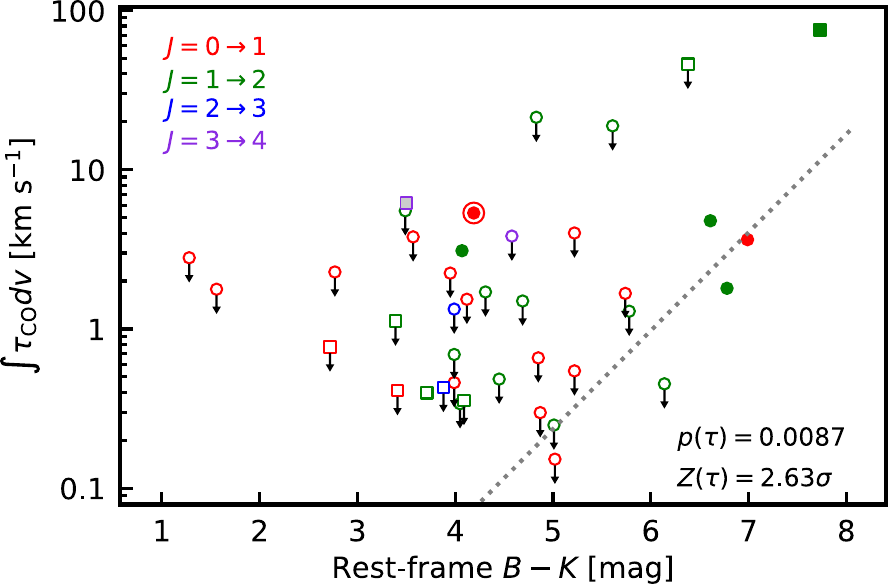}
\caption{The velocity-integrated optical depth of the CO absorption plotted against the rest-frame $B-K$ color of the sight-line. The associated systems are shown as circles and the intervening systems as squares. The circled red dot represents B2\,0902+34. The unfilled symbols/arrows show the upper limits, which were calculated based on the 3$\sigma$ limit across a single channel with a width of 10 \kms\ \citep[based on][]{cur11}. These upper limits are included in the linear fit, shown by the dotted line. The statistics to calculate the linear fit exclude 0201+113 (purple-gray square at $B-K$\,=\,3.5, $\int{\tau_{\rm CO}\delta v}$\,=\,6.3) for reasons clarified in the text.}
\label{fig:CO_B-K}
\end{figure}

\begin{deluxetable*}{lcccccccc}
\tablecaption{Existing millimetre-band CO detections. $z_{\rm abs}$ is the absorption redshift, followed by a flag designating whether this is associated with the continuum source (A) or arises in an intervening system (I). We then give the velocity-integrated optical depth of the CO absorption (in \kms), the rotational transition to which this applies, and the reference. The last three columns give the rest-frame $B$ and $K$ magnitudes as well as the resulting rest-frame $B-K$ color.}
\tablehead{
\colhead{Source} & \colhead{$z_{\rm abs}$} & \colhead{Type} & \colhead{$\int\tau_{\rm CO} \delta v$} & \colhead{Trans.} & \colhead{Ref.} &  \colhead{$B_{\rm rest}$} &  \colhead{$K_{\rm rest}$} &  \colhead{$B-K$} 
}
\startdata
0132--097 & 0.7634 & I & 75.6 &  $1\rightarrow2$ & W18 & 19.8 & 12.07 & 7.73\\
(0201+113  &  3.3872 &  I & 6.3  & $3\rightarrow4$ & C23 & 18.66 & 15.16 & 3.50)\\ 
0218+35   & 0.6885 & I &  20.3 &  $1\rightarrow2$ & W95 & 20.25 & --- & --- \\
0902+34   &  3.398 & A & 5.35 & $0\rightarrow1$ & this & 21.80 & 17.61 & 4.19\\
1200+045  & 1.2128 & A & 3.1 & $1\rightarrow2$ & C23 & 18.74 & 14.67 & 4.07\\
1245--19   & 1.2661 & A & 7.2 & $1\rightarrow2$ & C23 & --& 16.55& --\\
1413+135 & 0.2467 & A  & 3.6 & $0\rightarrow1$ & W97  & 21.01  & 14.02  &   6.99\\
1504+37   & 0.6734 & A & 12.2 & $1\rightarrow2$ & W96  & 21.44  & 14.66 &  6.78\\
1740--517  & 0.4423 & A  & 4.8 & $1\rightarrow2$ & A19 &  21.31  & 14.70  &   6.61\\
1830--211 & 0.8853 & I & 1.8 & $3\rightarrow4$ & W98 & 22.98 & --- & --- \\
\enddata
\tablecomments{References:  W95 -- \citet{wik95}, W96 -- \citet{wik96a}, W97 -- \citet{wik97}, W98 -- \citet{wik98}, W18 -- \citet{wik18}, A19 -- \citet{all19}, C23 -- \citet{com23}.}
\label{tab:mags}
\end{deluxetable*}

 \citet{cur06} noted that the sight-lines of damped Lyman-$\alpha$ absorption (DLA) systems, which have molecular fractions ${\cal F}\,\equiv\,\frac{2N_{\rm H_2}}{2N_{\rm H_2}+N_{\rm HI}}$\,$\sim\,10^{-7} - 0.3$, appeared to be much less reddened, with optical--near-infrared colors of $V-K\la4$, than the millimetre-band absorbers, which have ${\cal F}\approx0.6 - 1$ and $V-K\ga5$. That is, the  millimetre-band detections occur along much redder and optically fainter sight-lines, suggesting dustier, more molecular-rich gas. Thus, targeting sight-lines towards sources that are sufficiently unobscured in the optical band, which can yield a reliable redshift to which to tune the millimeter receivers, will lead to a bias against the detection of the cold, obscuring gas traced in the millimeter band \citep[see also][]{zwa06}. In fact, all but one of the previous CO absorption detections were follow-up observations of targets that had previously been detected in \HI\ 21-cm absorption \citep{uso91,car92,car98,che99,kan03a,kan03b,all19},\footnote{The exception is PKS\,1830--211, which was found through a 14 GHz wide spectral scan of the 3-mm band \citep{wik96b}.} which is also correlated with the $V-K$ color \citep{cur19}.

To test whether the absorption-line strength is consistent with the degree of reddening for 0902+34 and the other previous millimeter-band detections, Fig.~\ref{fig:CO_B-K} shows the distribution of integrated optical depth and rest-frame $B-K$ color for the known detections of CO absorption at $z_{\rm abs}$\,$>$\,0.1. The colors were obtained following the procedure described in Appendix\,\ref{sec:colors} and are listed for the detections in Table~\ref{tab:mags}. Included in Fig.\,\ref{fig:CO_B-K} are the upper limits from the unsuccessful searches for redshifted CO absorption, as summarized in \citet{cur11}, with the addition of those in \citet{com23}.

For the CO detections, there are only seven sources with the colors available. One of these is the intervening CO absorber towards PKS\,0201+113 (purple-gray square at $B-K$\,=\,3.5, $\int{\tau_{\rm CO}\delta v}$\,=\,6.3 in Fig.\,\ref{fig:CO_B-K}), which was detected at a tentative 3$\sigma$ level \citep[][]{com23}. 
Furthermore, this absorber occurs in a damped Ly$\alpha$ absorption system with a low molecular fraction ($6.3\times 10^{-7} \leq {\cal F} \leq 2.5\times10^{-5}$; \citealt{sri12})\footnote{As would be expected from its low
metalicity of [Z/H] $= -1.26$ \citep{cur04}.} and so it is very anomalous. We therefore exclude 0201+113 from our statistics. The remaining six CO absorbers do not show evidence of a correlation. However, when adding the limits of the non-detections, via the {\em Astronomy SURVival Analysis} ({\sc asurv}) package \citep{iso86}, a generalised non-parametric Kendall-tau test gives a probability of $P(\tau) =0.009$ of the distribution occurring by chance, which is significant at $Z(\tau) = 2.63\sigma$, assuming Gaussian statistics (see Table~\ref{tab:stats_table} of Appendix \ref{sec:statfit}). For this, the linear fit, including the limits, gives
\begin{equation}
{\rm log}_{10}\int\tau_{\text{CO}}\delta v \approx 0.62 (B-K) -3.72.
\label{eq:fit}
\end{equation}

Most of the published limits are in associated systems. Although four of the detected systems occur in absorbers intervening a more distant continuum source, the $B-K$ colour is only available for two of those, including 0201+113 (Table~\ref{tab:mags}). The other intervening absorber is 0132--097, which constitutes an ``end point'' that drives the correlation, with the largest absorption strength and degree of reddening (the green square at $B-K$\,=\,7.7, $\int{\tau_{\rm CO}\delta v}$\,=\,76 in Fig.~\ref{fig:CO_B-K}). Removing also this intervening system gives a significance of $1.71\sigma$. 

Figure\,\ref{fig:CO_B-K} shows that B2\,0902+34 appears to be an outlier, being much less reddened ($B-K$\,$\sim$\,4.2) than would be expected based on the linear fit. This is also the case for the CO absorber towards 1200+045 at $z$\,=\,1.2 \citep{com23}. However, the color of 0902+34 is quite uncertain (see Appendix A).\footnote{Removing B2\,0902+34, in addition to 0201+113, increases the significance of the correlation to $Z(\tau) = 2.90\sigma$ (see Table~\ref{tab:stats_table} of Appendix \ref{sec:statfit}).} Also, the different CO transitions shown in Fig.~\ref{fig:CO_B-K} likely trace molecular gas with different excitation conditions, which may introduce a scatter. Nevertheless, as more data are added, it would be interesting to see if the above fit, or any in Table \ref{tab:stats_table} (Appendix \ref{sec:statfit}), provides a useful diagnostic in predicting the CO absorption-line strength from the sight-line color, as can be expected if both properties depend on
the galaxy’s dust content.

\section{Conclusions}
\label{sec:conclusions}

We presented a multi-component molecular absorption system in the radio galaxy B2\,0902+34 at $z$\,=\,3.4, detected in \CO\ and \CN. This system was previously detected also in H\,{\sc I} 21-cm absorption of neutral gas. Summarizing the main results:
\begin{itemize}
\item{B2\,0902+34 is the highest redshift galaxy in which absorption of neutral and molecular gas has been detected.}
\item{The \CO\ absorption consists of two components, each tracing an H$_{2}$ column density of $N_{\rm H_2}$\,$\ga$\,3$\times$10$^{20}$ cm$^{-2}$. Combined with the large FHWM of the CO components (26 and 69 \kms), we discuss the scenario that the absorption likely originates from a collection of small (pc-scale) molecular clouds that are distributed across a region with also diffuse gas and high turbulence among the clouds.}
\item {The \CN\ absorption consists of three components, with CO/CN ratios of integrated optical depth ranging from $\la$0.4\,$-$\,2.4. This suggests that the molecular clouds are dense ($n_{\rm H_2}$\,$\ga$\,10$^{4}$ cm$^{-2}$) and likely irradiated by UV or X-ray emission, possibly near the central AGN or starburst.}
\item{Compared to other distant CO absorbers and the large number of CO non-detections in the literature, we find indications at the 2.6$\sigma$ level for a correlation between the absorption line strength and the rest-frame $B-K$ color, with $\log_{10}\int\tau_{\text{CO}}dv \approx 0.62 (B-K) - 3.72$. If confirmed, this could indicate that detectable CO absorption is more prevalent among galaxies with a higher dust content. Despite this, B2\,0902+34 shows less red colors than the CO absorption systems at z\,$<$1.}
\end{itemize}
\ \\
As a result of our VLA observations, the discovery of this multi-phase absorber, 1.5 Gyr after the Big Bang, opens a window for studying the neutral and molecular gas on small scales in distant galaxies during the era of the Square Kilometre Array (SKA; \citealt{laz09}) and Next-Generation Very Large Array (ngVLA; \citealt{mur18}).

\acknowledgments

The authors thank Ilse van Bemmel and Suma Murthy for for useful discussions. The National Radio Astronomy Observatory is a facility of the National Science Foundation operated under cooperative agreement by Associated Universities, Inc. Based on observations with the NASA/ESA Hubble Space Telescope obtained from the Data Archive at the Space Telescope Science Institute, which is operated by the Association of Universities for Research in Astronomy, Incorporated, under NASA contract NAS5-26555. Support for program number HST-GO-17268 was provided through a grant from the STScI under National Aeronautics and Space Administration (NASA) contract NAS5-26555. RJvW acknowledges support from the ERC Starting Grant ClusterWeb 804208. MVM acknowledges support from grant PID2021-124665NB-I00 by the Spanish Ministry of Science and Innovation (MCIN) / State Agency of Research (AEI) / 10.13039/501100011033 and by the European Regional Development Fund (ERDF) “A way of making Europe”.

%

\vspace{5mm}
\facilities{VLA, {\it HST}}


\software{CASA \citep{casa22}}

\appendix

\section{CO spectrum prior to continuum subtraction}
\label{sec:contsub}

\begin{figure}[h!]
\centering \includegraphics[width=0.5\columnwidth]{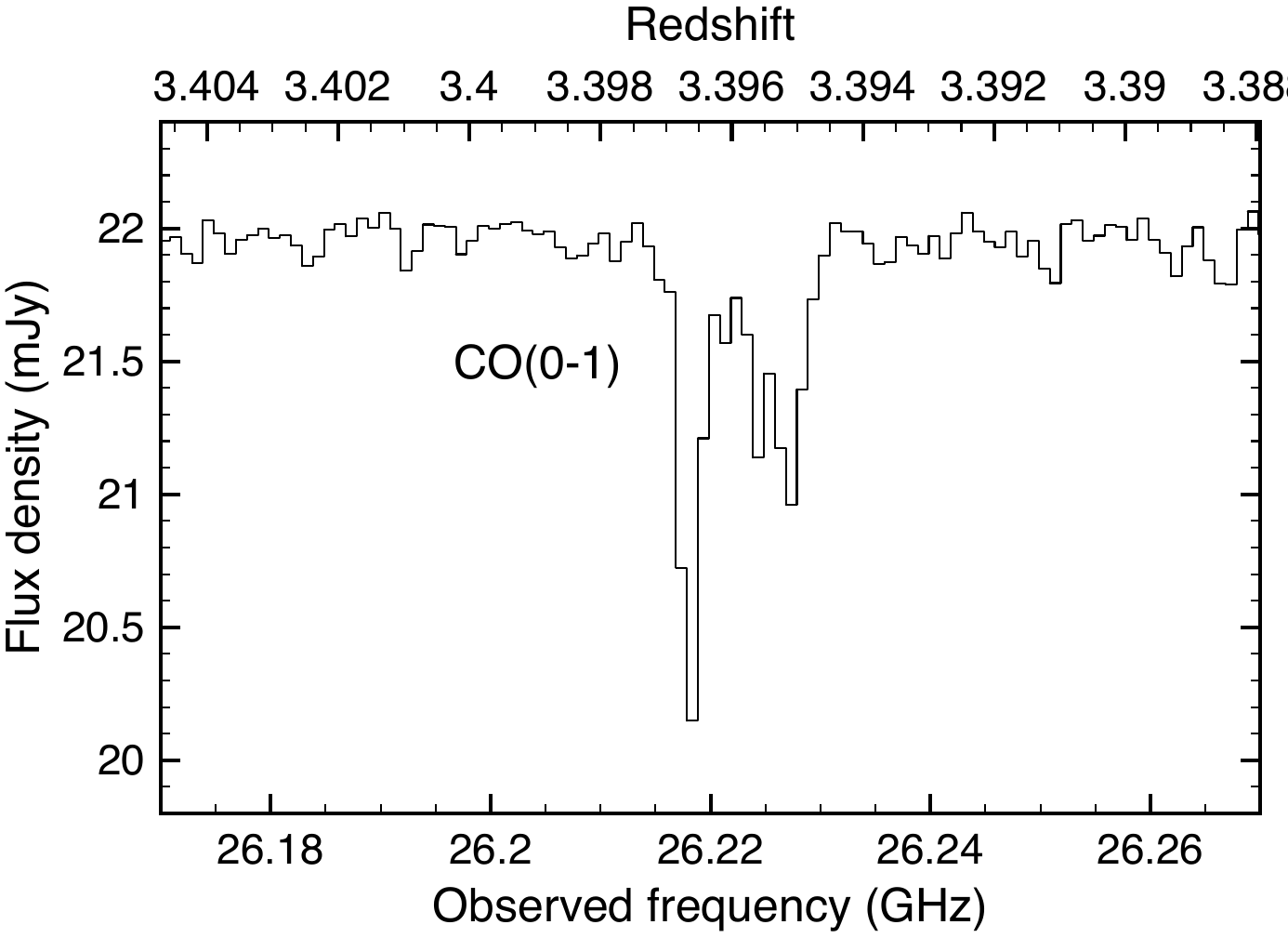}
\caption{\CO\ absorption spectrum superposed on top of the radio continuum. This spectrum was obtained from an image cube made before subtracting the continuum in the ($u$,$v$)-domain (Sect.\,\ref{sec:data}).}
\label{fig:linecont}
\end{figure}

Sect. \ref{sec:data} described why and how the radio continuum was subtracted in the ($v$,$v$)-domain prior to imaging the line data. Figure \ref{fig:linecont} shows the \CO\ absorption on top of the continuum in the image cube that was made prior to continuum subtraction. This illustrates that the continuum subtraction had no adverse effect on the absorption signal.

\section{Sight line colors to CO absorbers}
\label{sec:colors}

In order to test whether the absorption line strength of $\int\tau dv = 5.35$~\kms\ is consistent with the degree of reddening (Sect.~\ref{sec:reddening}), for B2\,0902+34 and the other previous detections and non-detections, we used the same method as \citet{cm19}, where we scraped the photometry from the {\em NASA/IPAC Extragalactic Database} (NED), the {\em Wide-Field Infrared Survey Explorer} (WISE, \citealt{wri10}) and the {\em Two Micron All Sky Survey} (2MASS, \citealt{skr06}).

\begin{figure*}
 \centering 
 \includegraphics[width=0.77\textwidth]{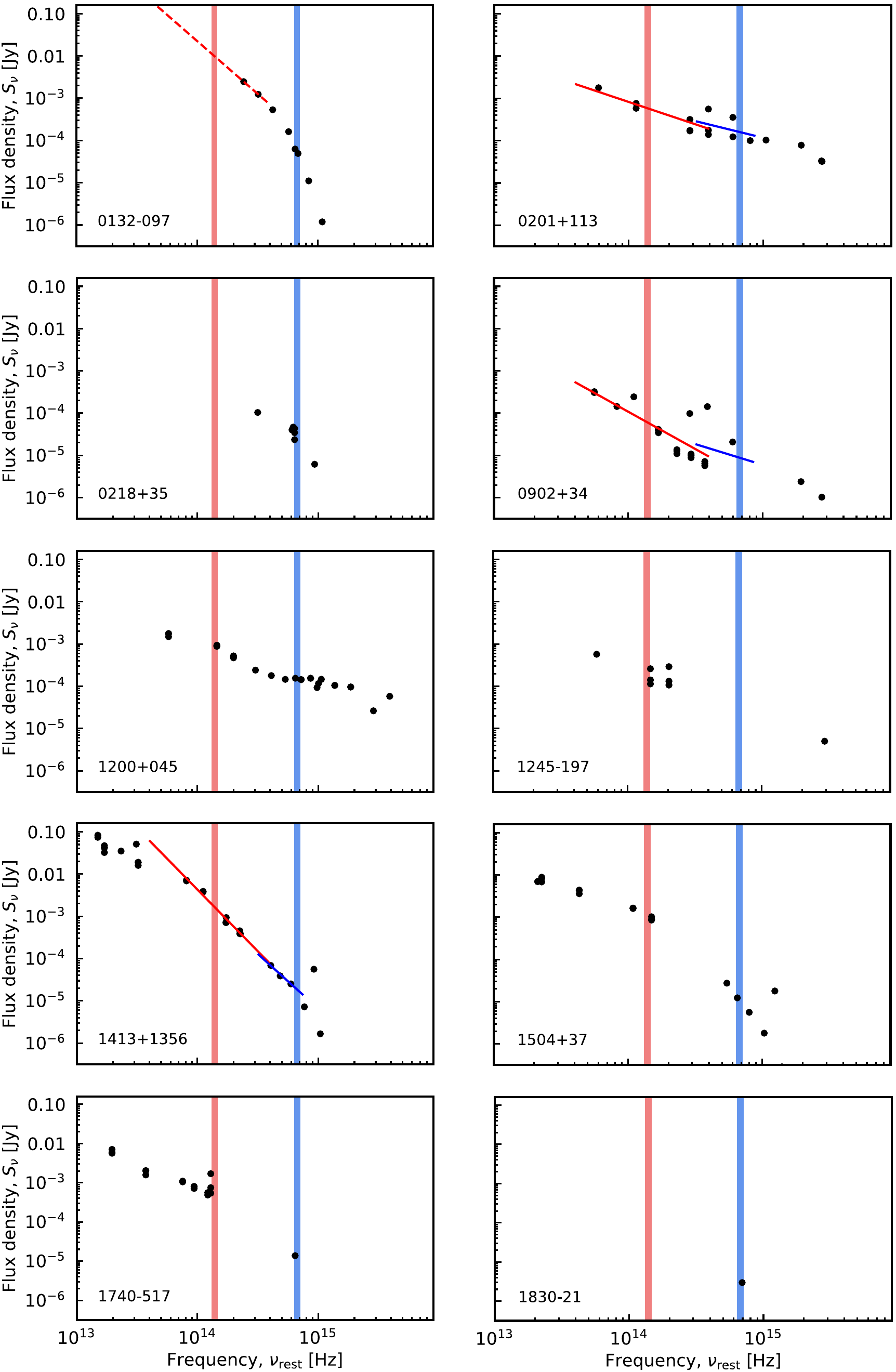}
 \caption{The optical and near-infrared photometry in the rest-frame of the absorber, for the sources detected in redshifted CO absorption. The lines show the power-law fits used to obtain the $K$ (red dashed) and $B$ magnitudes (blue dotted), where these did not fall into the bands. Results are summarized in Table \ref{tab:mags}.}
  \label{fig:SEDs}
\end{figure*}

Figure \ref{fig:SEDs} shows the broadband photometry from NED, WISE, and 2MASS of all known CO absorbers at $z$\,$>$\,0.1. Given the large redshift of B2\,0902+34, we shifted the SEDs back into the rest-frame. Following the same method as \citet{cur21}, if the frequency of the photometric point fell within $\Delta\log_{10}\nu=\pm0.05$ of the central frequency of the band, the flux measurement was added with multiple values being averaged. If no photometry for that band was available, this was obtained by a power law fit to the neighbouring photometric points. Results of the color analysis of the CO absorbers are summarized in Table \ref{tab:mags} and further details are provided in Sect.\,\ref{sec:reddening}.

\section{Statistics on correlation between absorption strength and $B-K$ colors}
\label{sec:statfit}

Table \ref{tab:stats_table} gives an overview of the statistics of the CO absorption strength versus rest-frame $B-K$ color for both the detections (Table \ref{tab:mags}) and non-detections \citep{cur11,com23}. This follows the analysis shown in Appendix \ref{sec:colors} and the equation
\begin{equation}
{\rm log}_{10}\int\tau_{\text{CO}}dv = m \cdot (B-K) + c,
\end{equation}
where $m$ is the gradient and $c$ is the intercept of the linear fit. See Table \ref{tab:stats_table} for more details.

\begin{table}
  \centering
  \begin{minipage}{73mm}
  \caption{Statistics of the CO absorption strength versus rest-frame $B-K$ color of the sight-line (Fig.~\ref{fig:CO_B-K}). These are given for the whole sample, the sample excluding 0201+113, the sample excluding 0201+113 and 0902+34, and the associated absorbers only. $n$ gives the sample size, $p(\tau)$ and $Z(\tau)$ the results of a Kendall-tau test and $m$ \& $c$ the gradient and intercept of the linear fit, respectively (see main text).}
    \begin{tabular}{@{}l ccccr @{}} 
    \hline
    \smallskip
    & $n$ & $p(\tau)$ & $Z(\tau)$& $m$ & $c$ \\
    \hline
    \multicolumn{6}{c}{\sc whole sample}\\
    Detections &  7 & 0.881 & $0.15\sigma$ & 0.10 & 0.81\\
    All &        40 &  0.050 & $1.96\sigma$ & 0.48 & $-2.89$ \\
    Associated only  & 31 & 0.098 & $1.66\sigma$ & 0.35 & $-2.22$\\
    \multicolumn{6}{c}{\sc excluding 0201+113} \\
    Detections & 6 & 0.573 & $0.56\sigma$ & 0.16 & $-0.19$\\
    All &        39  & 0.009 & $2.63\sigma$ & 0.62 &  $-3.72$\\
    Associated only  & 31 & 0.098 & $1.66\sigma$ & 0.35 & $-2.22$\\
    \multicolumn{6}{c}{\sc excluding 0201+113 and 0902+34} \\
    Detections &5 & 0.327 & $0.98\sigma$ & 0.23 & $-0.72$\\
    All &        38  & 0.004 & $2.90\sigma$ & 0.73 &  $-4.45$\\
    Associated only  & 30 & 0.033 & $2.14\sigma$ & 0.47 & $-2.94$\\
    \hline
    \end{tabular}
\label{tab:stats_table}
\end{minipage}
\end{table}

\end{document}